\documentclass[12pt]{article}
\usepackage{graphicx}
\newcommand{\sss}{\scriptscriptstyle}

\newcommand {\be}{\begin{equation}} 
\newcommand{\ee}{\end{equation}}    

\def\dds{\frac{\partial}{\partial s}}
\def\dds1{\frac{\partial}{\partial s_1}}

\def\vti{v_{{\sss T}i}}
\def\vte{v_{{\sss T}e}}
\def\vtj{v_{{\sss T}j}}
\def\vta{v_{{\sss T}a}}

\def\d{d\kern-0.8 ex\vrule height 1.3 ex depth-1.24 ex width 0.7 ex
\kern 0.15 ex}
\def\D{D\kern-1.7 ex\vrule height .87 ex depth-0.8 ex width 0.7 ex
\kern 0.95 ex}

\begin{document}

\baselineskip 18 pt
\begin{center}
  {\Large{\bf Collisions, magnetization, and transport coefficients  in  the lower solar atmosphere}}
\vspace{0.5cm}

{\bf   J. Vranjes}$^1$ and {\bf P. S. Krstic}$^2$\\
       $^1$Institute of Physics Belgrade, Pregrevica 118, 11080 Zemun, Serbia\\
 $^2$Joint Institute of Computational Sciences, University of Tennessee, Oak Ridge, TN 37831-6173, USA

\end{center}
 \thispagestyle{empty}
 
{\bf   Abstract:} The lower solar atmosphere is an intrinsically multi-component and collisional
environment with electron and proton collision frequencies in the range $10^{8}-10^{10}$ Hz,
which may be considerably higher than  the gyro-frequencies for
both species. Collisions between different species are altitude dependent
because of the  variation in  density and temperature of all species.
   
   We aim  to provide a reliable quantitative set of
data for  collision frequencies, magnetization, viscosity, and thermal
conductivity for the most important species in the lower solar atmosphere.
Having such data at hand is essential for any modeling that  is aimed at
describing realistic properties of the considered  environment.

The relevant elastic and charge transfer cross sections in the considered  range of
collision energies are now accepted by the scientific community as known with
unprecedented accuracy for the most important species that may be found in the lower solar
atmosphere. These were previously  calculated  using a quantum-mechanical approach and were validated by laboratory
measurements. Only with reliable
collision data one can obtain accurate values for collision frequencies and
coefficients of viscosity and thermal conductivity.
 
 We describe the altitude dependence of the parameters and the different physics of collisions
between charged species, and  between charged and neutrals species. Regions of dominance of each type of collisions are clearly
identified. We determine the layers within which either electrons or ions or both
 are unmagnetized. Protons are shown to be un-magnetized in the lower atmosphere in a layer that is at least
 1000 km thick even for a kilo-Gauss magnetic field that  decreases exponentially with altitude.   In these  layers
the dynamics of charged species cannot be affected by the magnetic field, and
this fact is used in our modeling. Viscosity and thermal conductivity
coefficients are calculated for layers where ions are unmagnetized.
We compare  viscosity and friction and determine the  regions of dominance
of each of the phenomena.

We provide the most reliable quantitative values for most important
parameters in the lower solar atmosphere to be used in analytical modeling and
numerical simulations of various phenomena such as waves, transport and magnetization of particles,
and the triggering mechanism of coronal mass ejections.
 
\section{Introduction}
There has recently been  a considerable shift of focus of solar researchers
from ideal and collision-less  toward collisional phenomena in the solar atmosphere (e.g., Arber et al. \cite{arb},
 Pandey and Wardle \cite{pan}, Soler et al. \cite{sol}, Barcel\'{o} et al. \cite{bar},  Zaqarashvili et al. \cite{zak2},
 \cite{zak},  Khomenko and Collados \cite{kho}). This is understandable for lower atmosphere layers because this
is a partially ionized medium with several species whose dynamics is heavily
coupled  and affected by mutual collisions and by collisions
between similar particles (Vranjes and Poedts \cite{v1}, Vranjes et al. \cite{v2, v3}). In some layers of the photosphere the ion-neutral and
electron-neutral collision frequencies are approximately $10^{9}$ and $10^{10}$ Hz, respectively. This makes both ions and electrons in these  layers very weakly magnetized or completely un-magnetized.

In recent multi-component models in the literature, the medium has typically
been treated as if it consisted of  two components,  neutrals and `plasma'. These  models
consequently neglected differences between electron and ion dynamics,
similarly to ordinary magneto-hydrodynamics theory. However, in the lower
solar atmosphere it is very difficult to justify this approach, as we show
below. One reason for this are the different collision frequencies of electrons
and ions, which among other effects imply different magnetization of these
two species, consequently producing quite different effects of the magnetic field on particle dynamics.

The key components in the lower solar atmosphere are identified in the
present work. We also show their collision cross sections and collision
frequencies as a function of altitude. Using these results, the altitude-dependent magnetization of electrons and ions, the viscosity tensor, and
the thermal conductivity vector components are calculated. The altitude-dependence of the parameters is presented graphically.

\section{Key ingredients}

The lower solar atmosphere is an essentially multi-component system that
includes electrons, protons, neutral hydrogen H and helium He atoms, helium
ions He$^{+}$ or/and He$^{++}$ etc. The dissociation energy of  molecular hydrogen H$_{2}$  is  $\simeq 4.5$ eV and its  presence may be expected throughout the lower solar
atmosphere. However,
recent observations (Jaeggli et al. \cite{jaeg}) showed that the  regions with the highest  amount of molecular hydrogen are sunspots  where the total fraction of $H_2$ molecules may reach a few percent only. On the other hand, the ionization energy of the hydrogen molecule, $15.603$ eV, exceeds the ionization energy of atomic hydrogen, so the presence of {\em the ionized} hydrogen molecule is  most likely negligible.
 Having in mind all possible collisional combinations of various particles, it is essential to
identify the most important of their interactions before making any attempt
to obtain  realistic models for wave damping, diffusion, transport,
magnetization of particles, etc. The collision cross sections can vary
with the energy of colliding particles, which in the solar atmosphere is
equivalent to the variation of the temperature with altitude.

Detailed studies available in the literature (Krstic and Schultz \cite{kr3},
Glassgold et al. \cite{kr1}, Schultz and Krstic \cite{kr2}, Jamieson et al. \cite{jam} etc.) show
large differences between the cross sections describing collisional
phenomena such as elastic scattering, momentum transfer, viscosity, spin
exchange, and charge exchange. Among these the most prevalent is typically the elastic scattering. Physically, this cross section should be taken
into account in estimating the magnetization. On the other hand, the cross
section for momentum transfer should be used in calculating  friction;
this type of cross section turns out to be usually lower than the elastic scattering.
Knowing  these details may be essential in realistic modeling of the
lower solar atmosphere.

One also needs to include  some inelastic collisions, which clearly may play a decisive
role in the partially ionized solar atmosphere (Vranjes and Poedts \cite{vpla}), which
makes the whole analysis significantly more complicated. These
processes are the radiative recombination (of the type $A^{+}+e\rightarrow
A+h\nu $), three-body recombination (of the type $A^{+}+e+e\rightarrow A+e$),
dissociative recombination (e.g. of the type $A_{2}^{+}+e\rightarrow
A+A^{\ast }$, where $A^{\ast }$ is an excited atom), ionization by electron
impact (the process of the type $A+e\rightarrow A^{+}+e+e$), the ion-atom
(or molecule) interchange reactions of the type $A^{+}+BC\rightarrow
(AB)^{+}+C$, etc.
\begin{figure}[tbh]
\centering
\includegraphics[height=6cm,bb=16 16 272 222,clip=]{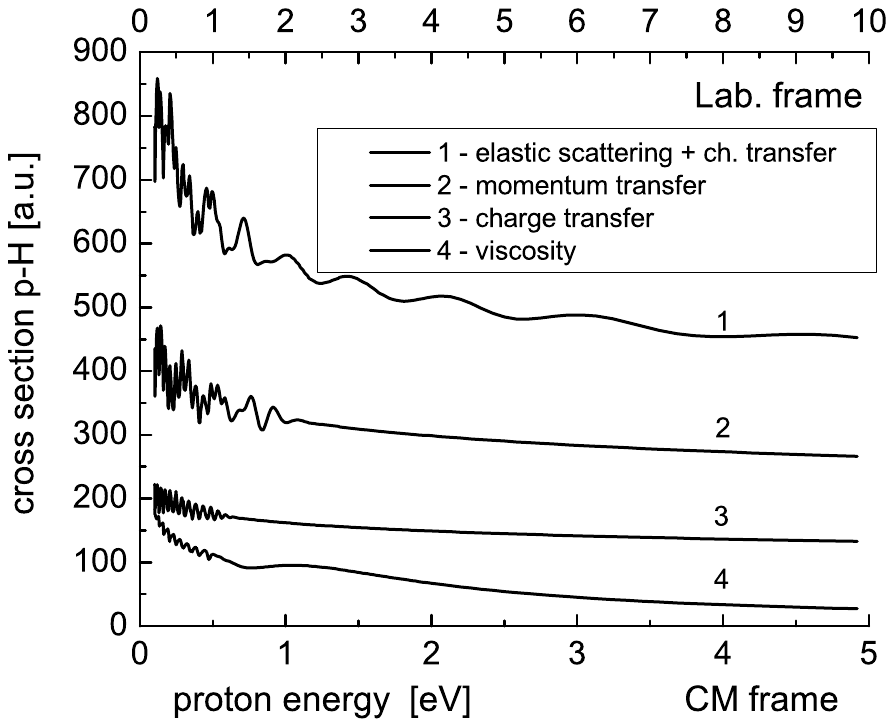}
\caption{Integral cross sections $\protect\sigma _{p{\scriptscriptstyle H}}$
for proton ($p$) collisions with neutral hydrogen atoms H for quantum-mechanically indistinguishable nuclei of the projectile and target particles, following
Krstic and Schultz \cite{kr3}. Here and throughout  the text 1 a.u.
$=2.8\cdot 10^{-21}$ m$^{2}$, $1\,eV\simeq 11604$ K.}
\label{fig1}
\end{figure}
The presence of inelastic collisions implies modifications
of equations through appropriate source/sink terms that appear in the
continuity equation, in the momentum and in the energy equations.
Only collisional phenomena from the first group mentioned above, i.e.,
elastic and charge transfer  processes,  are  discussed here. Even then, as
will become evident in the following sections, we still encounter a
plethora of processes that could  further be classified by priority and
dominance  to perform any meaningful  analysis.
\begin{figure}[tbh]
\centering
\includegraphics[height=6cm,bb=17 14 265 222,clip=]{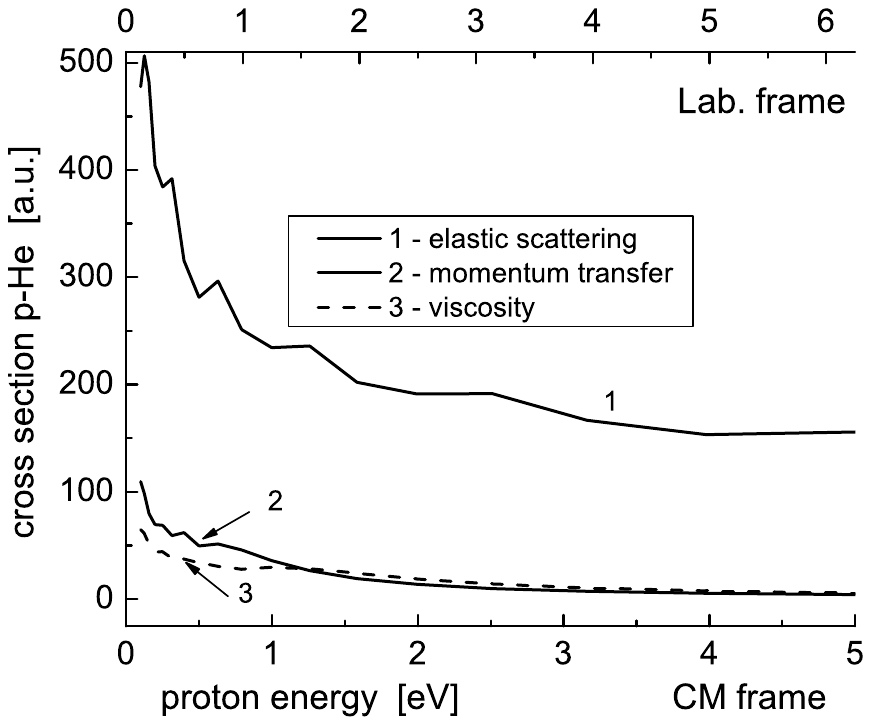}
\caption{Integral cross sections $\protect\sigma_{p{\scriptscriptstyle H}e}$
for proton collisions with neutral helium He. }
\label{fig2}
\end{figure}
There exists a numerous literature dealing with collisional plasmas in the
laboratory environment. The laboratory knowledge is in principle directly
applicable to the relevant phenomena in the solar atmosphere. However, we
find that i) the theoretical and experimental results are  scattered to a
large extent, and, ii) for the purpose of  solar plasma studies, the
cross sections for a specific range of parameters of the solar plasma are
missing. For this reason  we present cross
sections for possible key-players in the lower solar atmosphere in the following subsection, with the
energies of colliding particles in the range appropriate for the plasma
below the transition region. These results, combined with the existing
models of the altitude-dependent densities and temperature of the plasma
species, are the essential and sufficient ingredients for calculating
collision frequencies that are necessary for modeling.

 \subsection{Cross sections for charge-neutral and neutral-neutral collisions}\label{cn}

\begin{figure}[tbh]
\centering
\includegraphics[height=6cm,bb=17 16 266 223,clip=]{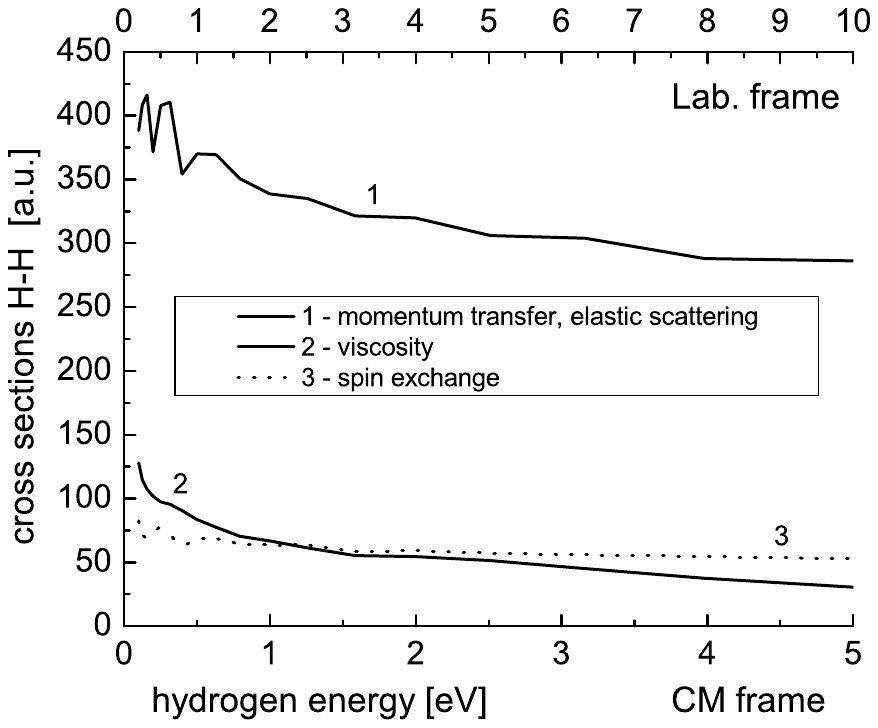}
\caption{Integral cross section $\protect\sigma_{{\scriptscriptstyle HH}}$
for collisions between neutral hydrogen atoms H for quantum-mechanically indistinguishable nuclei. }
\label{fig3}
\end{figure}

\begin{figure}[tbh]
\centering
\includegraphics[height=6cm,bb=16 15 280 212,clip=]{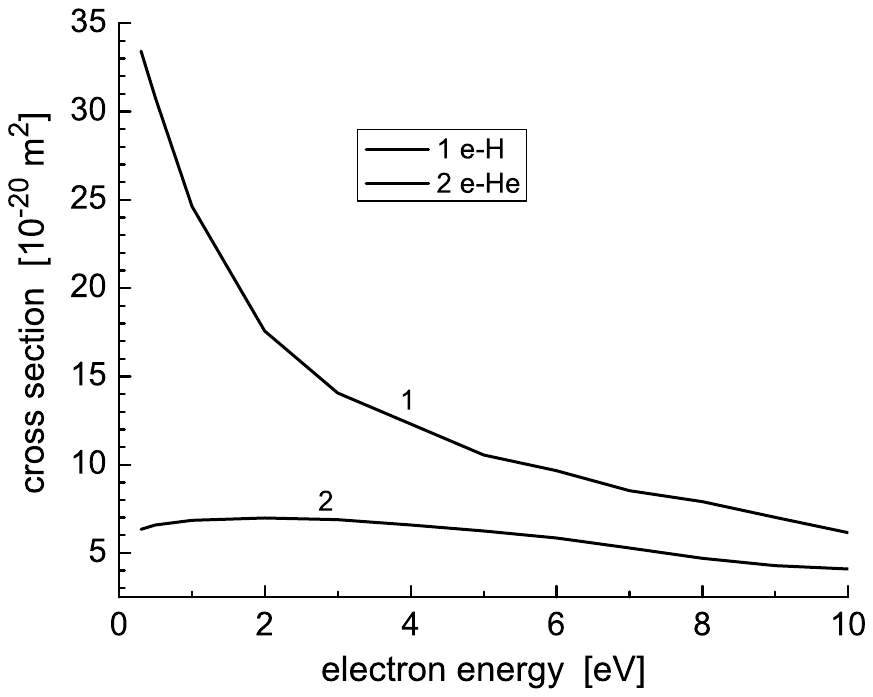}
\caption{Cross section for elastic scattering of electrons on neutral
hydrogen atoms H and neutral helium
atoms He, in terms of the electron energy. }
\label{fig4}
\end{figure}

Figs.~\ref{fig1}-\ref{fig4} described in this section provide cross sections
for all most important ingredient species in the lower atmosphere, given in
terms of the energy (temperature) of the colliding species, which makes them
directly applicable to the varying temperature with the altitude. The cross section
profiles in Figs.~\ref{fig1}-\ref{fig3} are based on works of Krstic and
Schultz \cite{kr3}, \cite{kr4}, \cite{kr44}. These authors derived both differential and integral cross
sections for elastic scattering.  The cross sections for the
momentum transfer and viscosity are the first and second
moments of these [see also Dalgarno et
al. \cite{dal2},  Hasted \cite{hast},  Brandsen \cite{brand},  Makabe and Petrovic \cite{map}, Schunk and  Nagy \cite{snag}].
In the first (momentum transfer) the differential elastic
cross section is weighted by a scattering angle $\theta$ dependent term,
$1-\cos \theta $, while in the second (viscosity) it is weighted by
$\sin^{2}\theta $. These weighting factors describe different features associated with momentum transfer and viscosity. For the viscosity this emphasizes the scattering at an  angle $\pi/2$ and de-emphasizes the forward and backward  ones. Scattering to such large angles is very effective in equalizing energies of the colliding particles. This is seen from the expressions for energies of the two particles before ($E_1$, $E_2$) and after collisions ($E_1^{'}$, $E_2^{'}$) in the laboratory frame,  expressed through their total energy $E$ in the laboratory frame and  scattering angle $\theta$ in the CM frame: $E_1^{'}=E (1+\cos \theta)/2$, $E_2^{'}=E (1-\cos \theta)/2$. Hence, such large-angle scattering tends to  reduce  both  viscosity and conductivity.
 On the other hand,  the factor $1-\cos \theta$ in the momentum transfer cross section
emphasizes the backward-scattering angles, and this cross section determines the average momentum lost in collisions. Note also that charge exchange is a backward-scattering process. Many more details on these cross sections can be found  in  Krstic and Schultz \cite{kr3}, \cite{kr44}, currently  the
most accurate  cross sections for elastic processes and resonant charge
transfer.  The energy range in Figs.~\ref{fig1}-\ref{fig3}  in the center of mass (CM) of colliding particles is $0.1-5$ eV (bottom $x$-axis), while in the laboratory (plasma) frame the energy range is given by the top $x$-axis using the transformation formula $E_{lab}= E_{\sss {CM}} (m_1 + m_2)/m_2$, where $m_2$ is the mass of the target particle,  and 1 a.u. $=2.8\cdot 10^{-21}$ m$^{2}$, $1\,eV\simeq 11604$ K.

We start with the cross sections for proton collisions with neutral hydrogen
(p-H), shown in Fig.~\ref{fig1}. They are based on quantum-mechanical indistinguishability of the projectile and target nuclei
(Krstic and Schultz \cite{kr3}). The elastic scattering curve from Fig.~\ref{fig1}
(line 1) was  used to  calculate magnetization, i.e., for the ratio of the
collision frequency and the gyro-frequency of protons. It is the sum of the pure elastic scattering cross section and the
charge transfer cross section.  In estimating the
friction caused by neutral hydrogen, it is appropriate to use the momentum
transfer cross section (line 2). The amplitude oscillations of the
cross sections are the  consequence of quantum effects, which are present only
at lowest collision energies  (in the present study this means throughout the photosphere and chromosphere).

We stress that when one approximates the distinguishable particles,
the elastic scattering cross sections and their higher momenta (momentum
transfer and viscosity) are lower. This is because in this case we assume  that we can
distinguish between elastically and charge-transfer scattered protons, resulting in a
separate treatment of these processes. When these particles cannot be
distinguished, as is the case at lowest energies (lower than 1 eV), the elastic cross sections of the
indistinguishable particles  and their moments are the
coherent sum of  the processes, elastic and charge transfer. For indistinguishable nuclei, the
integral elastic cross section together with the charge transfer at $0.2, 0.5, 1$ eV in the CM frame is,
$788.660, 679.534$, and $ 582.292 $ a.u., while the distinguishable nuclei model yields $598.068, 506.701$, and $ 419.739$  a.u.
 Similar holds for the momentum transfer, while the
charge transfer cross section is practically the same for both models. One
has to have in mind these differences and  the differences in the
physical definitions of these cross sections, to avoid twice counting
 elastic and charge transfer cross sections: the elastic  cross section of the  indistinguishable
particle   and their moments in Fig.~\ref{fig1} already contain both
processes.

In Fig.~\ref{fig2} the three lines describe the collisions between protons
and neutral helium atoms. Observe that the momentum transfer and viscosity
lines are  below the line for elastic scattering by a factor 4-5. This is because the  momentum transfer presents the differential cross
sections in the backward-scattering directions, while the dominant
contribution in the elastic cross sections comes form the forward-scattering
angles, which dominate the differential elastic cross sections. Therefore,
for the proton dynamics the presence of neutral helium may be more important
for estimating magnetization than for the momentum loss caused by friction or by  viscosity (as compared
to those that come from proton self-collisions or interaction with hydrogen, see more in Sec. \ref{visg}).

Fig.~\ref{fig3}  gives  collision cross sections
between neutral hydrogen atoms according to Krstic and Shultz \cite{kr3}. The momentum transfer and elastic collisions curves coincide, and line 1 in the two  figures describes the most dominant interaction for the solar
atmosphere.  Note that it includes  both direct and recoil scattering as a direct consequence of the indistinguishability of particles,
and the same holds for the viscosity cross section. Due to these reasons the presented values are twice as high as the classic values obtained from the model of distinguishable particles.

The new type of cross section  that appears in Fig.~\ref{fig3} is
the spin exchange cross section, which describes collisions  in which  electrons (from the two colliding atoms) with different spin orientation  are exchanged. Processes of this type are the reason for the  cooling phenomena in the upper atmosphere, interplanetary space, and galactic HI regions (Bates \cite{bat}, Purcell and Field \cite{purc}, Dalgarno \cite{dal}).

Observe the difference between the lines 1 in Fig.~\ref{fig1} and in Fig.~\ref{fig3}. One
obvious reason for this  is the charge transfer cross section, which is contained in  line 1 in Fig.~\ref{fig1}.
 This can be subtracted to obtain the pure elastic scattering value $\sigma_{el, p{\sss H}}$. The cross section obtained in this way, at high energies, tends smoothly   toward  the corresponding elastic cross section obtained from the classic model of distinguishable particles. At the low  energies of interest here,  0.5 and 1 eV (in CM frame),  this yields  $\sigma_{el, p{\sss H}}=507.333, 420.038$ a.u. as the pure elastic scattering cross section for p-H collisions.  For the H-H collisions in Fig.~\ref{fig3} the corresponding values are lower by about   $137$ and $81$ a.u. for the two energies, respectively. The difference that  still remains (i.e., greater p-H than H-H cross section) should be attributed to  the fact that proton collision with H atoms causes charge polarization on the neutral atom [see  Chen and Chang \cite{cc2} and in Vranjes et al. \cite{vpre}].
This means that  to some extent p-H collisions involve features of the Coulomb interaction.  The physics behind this is as follows: a point charge $q_{0}$ placed at some distance from an atom with the radius $a$ that has a
point positive charge $q$ in the core and uniform negative cloud $-q$ around
it, will cause displacement of the initially uniform cloud charge. If the
external point charge $q_{0}$ is an electron, this displacement is in the
direction away from the electron position. If the external point charge $%
q_{0}$ is a positive ion, this displacement of the cloud charge will be
toward the ion and the force will again be attractive. The energy from
this attractive Coulomb interaction according to Chen and Chang \cite{cc2}
is
\begin{equation}
E_{att}=-\frac{q_{0}^{2}a^{3}}{8\pi \varepsilon _{0}r^{4}}.  \label{p1}
\end{equation}%
Hence, collisions of charged species with neutrals indeed involve a
Coulomb-type interaction, which affects collisions at very short distances,
see also  Dalgarno et al. \cite{dal2} and  McDowell and  Coleman \cite{mcd}.

Finally, we give plot in Fig.~\ref{fig4} with three lines for the cross sections
for the \emph{electron} scattering on the two most important atoms hydrogen H and  neutral helium atoms He
 in the electron energy range $0.1-10$ eV.
The lines  represent some mean values
from many references, see for example  Bedersen and Kieffer \cite{bk}
and references cited therein. The possible uncertainty is almost  10-25
percent  at low energies.
This also agrees  with some other sources, for instance  Brode \cite{brode},
Brackmann et al. \cite{brak}, Kieffer \cite{kief}, Mitchner and
Kruger \cite{mich}, Tawara et al. \cite{taw}, and Fortov et al. \cite{fort}. Hence, although some
uncertainty for electron cross sections exists, it is not substantial.
Clearly, at the low energies of interest for photosphere and chromosphere the
most probable are electron collisions with atomic hydrogen H. We investigate  the
electron Coulomb collisions in  the following section.

\section{Electron collision frequencies}

  To describe  the collisions between charged particles we use the following expression for the collision frequency between the charged species $b$ and $a$ following Spitzer \cite{spit} and  Vranjes et al. \cite{v33}:
    \be
        \nu_{ba}= 4\left(\frac{2\pi}{m_b}\right)^{1/2}\left(\frac{q_aq_b}{4 \pi \varepsilon_0}\right)^2 \frac{n_a L_{ba}}{3(\kappa T_b + \kappa T_a m_b/m_a)^{3/2}}, \label{e1}
        \ee
        \[
               L_{ba}=\log[r_d/b_0],\quad
                r_d=\frac{r_{da} r_{db}}{(r_{da}^2 + r_{db}^2)^{1/2}}, \quad r_{dj}= \frac{v_{{\sss T} j}}{\omega_{pj}},
                \]
                \[
    b_0=\frac{|q_a q_b|/(4 \pi\varepsilon_0)}{3\kappa (T_a+ T_b)}, \quad v_{{\sss T} j}^2=\frac{\kappa T_j}{m_j}, \quad \omega_{pj}^2=\frac{q_j^2 n_j}{\varepsilon_0 m_j}.
                           \]

As  is well known, the Coulomb logarithm $L_{ba}$ (introduced by Spitzer) describes the cumulative effect of numerous small angle deflections that are intrinsic to Coulomb-type collisions.

\begin{figure}[!htb]
   \centering
   \includegraphics[height=6cm,bb=16 15 276 211,clip=]{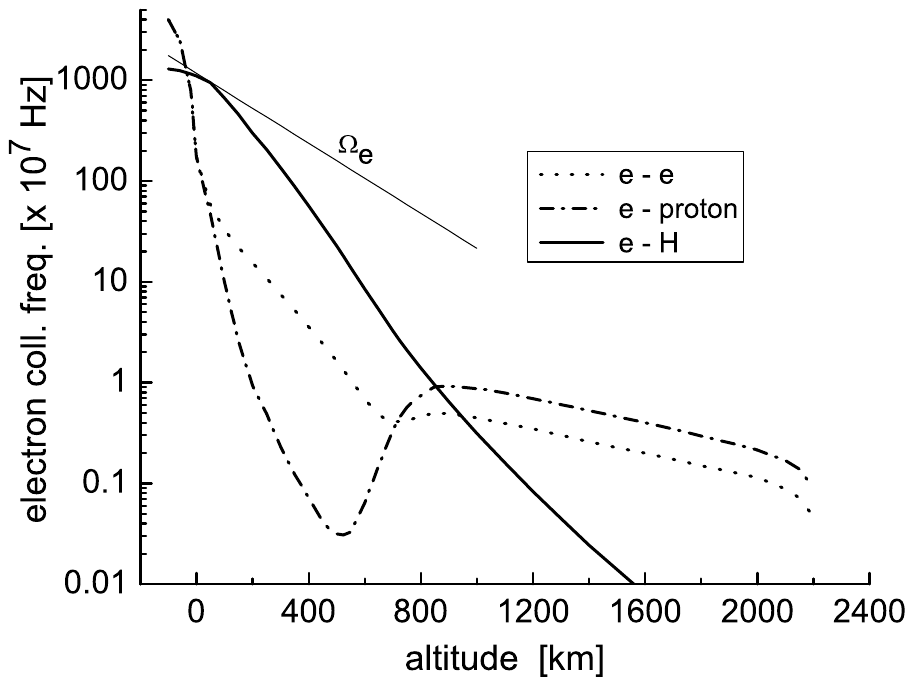}
\caption{Electron collisions with the altitude. For comparison,  the thin  line gives the electron gyro-frequency $\Omega_e$  for the starting   value of the magnetic field $B_0=0.1$ T. \label{fig5}}
\end{figure}

These expressions are  used  to calculate collisions  for electrons for the parameters (density and temperature) that vary with  altitude, The results are presented in Fig.~\ref{fig5}. To incorporate  the variation of the parameters, here and throughout  the text we use the values given in Table C in Fontenla et al. \cite{fon}.

  For electron collisions with neutrals we  read the cross sections $\sigma(x)$ from Fig.~\ref{fig4},  and then calculate the collision frequency using the expression
\be
\nu_{ej}=\sigma_{ej}(x) n_{j}(x) \vte(x). \label{nej}
 \ee
 The small differences  of the parameters (density and temperature) given in the reference above,  as compared with  some other models of the lower solar atmosphere from the same authors or others, are of no particular importance for the general picture that is  obtained.

The same holds for the expression $\nu_{ej}$ used here in comparison with some modifications of it that may be seen in the literature. For example, the thermal speed we use is without any numerical parameter, as for  the mean velocity $v=[8 \kappa T/(\pi m_{ab})]^{1/2}$, $m_{ab}=m_a m_b/(m_a + m_b)$, which is sometimes used in the literature. It is easily seen that  in the most drastic  case, for example  when $a=b$, this increases our thermal speed  by a factor 2.2 only. Similar numerical parameters appear in the most probable speed $v=(2 \kappa T/m)^{1/2}$, and in the root-mean-square velocity $(\hat{v}^2)^{1/2}=(3 \kappa T/m)^{1/2}$.  However, these  modifications are not substantial in view of our much more accurate cross sections  as compared with those that are typically used in the solar plasma literature (see comments in Sec. \ref{sum}).
\begin{figure}[!htb]
   \centering
\includegraphics[height=6cm,bb=17 14 272 215,clip=]{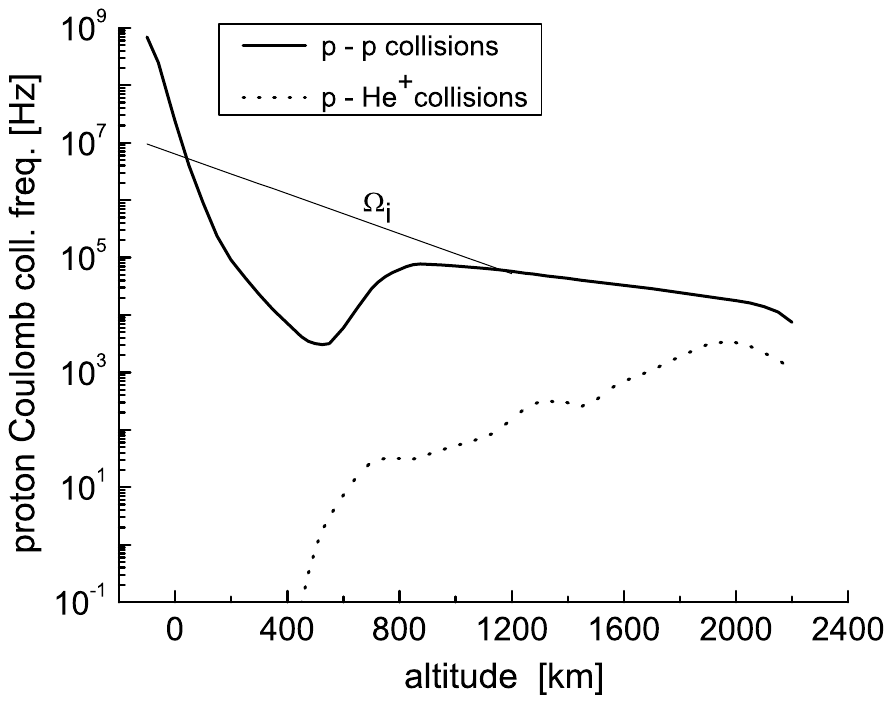}
\caption{Coulomb collisions of protons in terms of  altitude. \label{fig6}}
\end{figure}

 The electron collision frequency was checked also for e-He$^+$ collisions. For the given altitude range the maximum e-He$^+$ collision frequency is at about  $x=2000$ km altitude, but  it is only $0.02$ [in the same units as in Fig.~\ref{fig5}] and  is therefore completely negligible.
 After checking for the electron collision frequency with He$^{++}$ ions we found  out that it was  even lower,  at least by one order of magnitude.

 The   electron collisions with neutral helium He are even lower than  the dominant collisions in Fig.~\ref{fig5}. For example, at $x=0$ we have $\nu_{e{\sss H}e}=2.2\cdot 10^8$ Hz, which is almost two orders below  $\nu_{e{\sss H}}$, and it remains well below in the whole region.

 Fig.~\ref{fig5} suggests that in the region  $0-850$ km the electrons' collisions with atomic hydrogen are by  at least two orders of magnitude more frequent than the electron-electron collisions. Above 850 km e-p  collisions (and the associated  friction) should be  more important than both e-e collisions and electron collisions  with  neutrals. Below the level denoted by $x=0$ the Coulomb collisions become more dominant.
    \begin{figure}[!htb]
   \centering
   \includegraphics[height=6cm,bb=16 15 270 211,clip=]{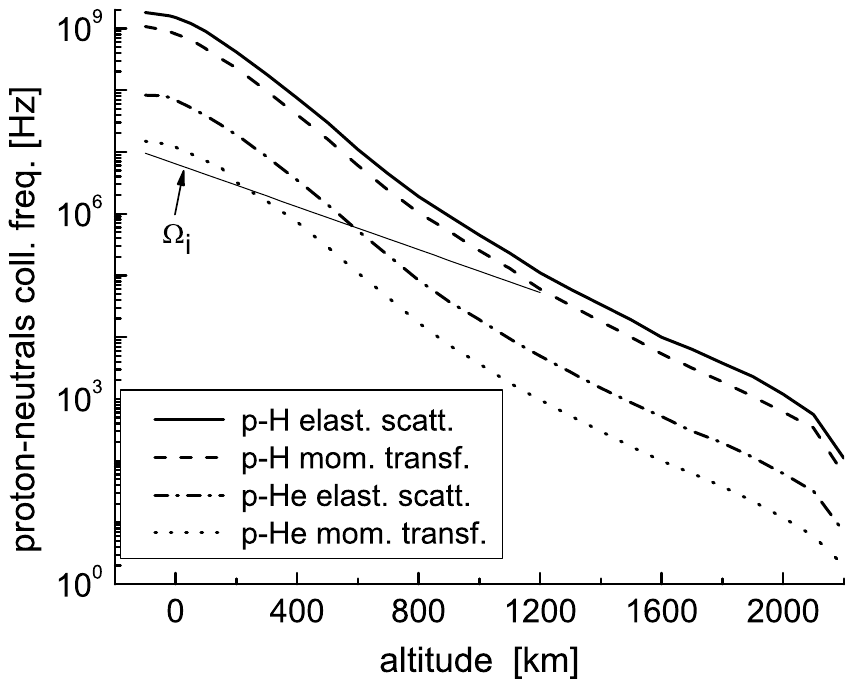}
\caption{Collision frequency for protons  colliding with the neutral atoms H and  He.  \label{fig7}}
\end{figure}

 According to  Table C in Fontenla et al. \cite{fon}, at 850 km  altitude the proton and H number densities are about  $10^{17}$ m$^{-3}$ and $ 10^{20}$ m$^{-3}$, respectively, i.e.,  neutral hydrogen atoms are about  1000 times more abundant, yet e-p collisions are already more frequent. A similar situation is  observed in the interval between $x=-100$ km  and   $x=0$.   Here and throughout  the negative altitude denotes the value below the referent level $x=0$, i.e., toward the center of the Sun. This all confirms the well-known fact that the Coulomb collisions have a much larger cross section and are more frequent even in rather weakly ionized plasmas  (see in Ratcliffe \cite{fra}, Vranjes and Poedts \cite{v08}). These facts are frequently overlooked in the literature.

\section{Proton collision frequencies}\label{procol}

Proton collision frequencies  $\nu_{pj}$ where $j$ includes protons as well as  other relevant charged or neutral  species are presented in Figs.~\ref{fig6} and \ref{fig7}.   Here, for proton-neutral collisions we have $\nu_{pj}= \sigma_{pj} n_j \vti$ and $  \sigma_{pj}$ is given in Figs.~\ref{fig1} and \ref{fig2} in lines 1 and 2.   In Fig.~\ref{fig6} the Coulomb p-p and p-He$^+$ collision frequencies are presented using Eq.~(\ref{e1}). Although p-He$^+$ collisions are obviously much less frequent, their actual importance may be understood only by comparing the friction and viscosity terms in the momentum equation, see Sec. \ref{visg}.

In Fig.~\ref{fig7}, the cross section $\sigma(x)=\sigma(T(x))$  is obtained from Figs.~\ref{fig1} and \ref{fig2} for the energies in the plasma frame (i.e., those given by the top $x$-axes), and the density and temperature (energy) from  Table C in  Fontenla et al. \cite{fon}.

 Comparing Figs.~\ref{fig6} and \ref{fig7}, clearly  proton collisions with neutral hydrogen are by far the most dominant
in the given range (which starts from $x=-100$ km) up to the altitude of about  1350 km (there is a difference of almost four  orders of magnitude  for collisions  in certain lower layers). Above 1350 km the amount of neutrals is sufficiently reduced  so that p-p collisions become dominant. The   highest proton collision frequencies  at $x=-100$ km read $\nu_{p{\sss H}}=2\cdot 10^9$ Hz, $\nu_{pp}=7\cdot 10^8$ Hz, $\nu_{p{\sss H}e}=8\cdot 10^7$ Hz.

Above 1900 km proton collisions with He$^+$ are more frequent than p-H, which  is even more true  for p-He collisions.  For example, at $x=2017$ km we have $\nu_{p{\sss H}e^+}=3.3\cdot 10^3$ Hz [see Fig.~\ref{fig6}], while  $\nu_{p{\sss H}}=1.2\cdot 10^3$ Hz and $\nu_{p{\sss H}e}=62$ Hz [see Fig.~\ref{fig7}],  and higher up this  difference increases. Accordingly,  above this layer  proton {\em friction} with (any) neutral atoms is negligible.

\section{Electron and proton magnetization}

To estimate the magnetization,  the   thin  line in Fig.~\ref{fig5} gives the electron gyro-frequency $\Omega_e(x)= e B_0(x)/m_e$ for a starting value of the magnetic field $B_0=0.1$ T which approximately changes exponentially  with the altitude as $\exp[-x/(2 h)]$, $h=125$ km,  following the thin-flux tube model and the usual pressure balance conditions. In the region below $x=0$ the Coulomb collisions (e-p and e-e) become more dominant [see Fig.~\ref{fig5}],  and clearly the collision frequency in that area is higher than the electron gyro-frequency even for  $B_0=0.1$ T.  Hence, in this  layer $\nu_e/\Omega_e>1$ and  electron  dynamics should not be influenced by the magnetic field (see also Vranjes et al. \cite{v2}).

 The layer without magnetization is  much wider  for protons and other ions. The corresponding line for protons is given in  Figs.~\ref{fig6} and \ref{fig7}. For the same  exceptionally strong  starting field $B_0=0.1$ T we have    $\Omega_i\simeq 9.6$ MHz, which  changes with altitude in such a way that it remains below $\nu_{p{\sss H}}$ up to at least 1000-1200 km. Above this altitude  the magnetic canopy is formed and the field changes less rapidly; it is therefore expected that above this altitude the profile for  $\Omega_i(x)$ is less steep.  From Figs.~\ref{fig6} and \ref{fig7} it is seen that ions remain  un-magnetized within  a layer of unknown width for $x<0$. Accordingly,  because no much stronger  magnetic field can be expected elsewhere in photosphere, we can conclude that there exists a layer {\em throughout the photosphere} which is  at least 1000 km  thick (most likely it is even thicker) within which protons remain un-magnetized in absolute sense.

 Geometry and magnitude of the magnetic field vary both horizontally and vertically. Therefore the  width of the layer within which protons are unmagnetized might be  expected to be much  wider in regions with a considerably  weaker field. However, assuming that the magnetic canopy forms at about  the altitude of 1000 km (c.f. Khomenko et al. \cite{el}), this implies that protons are unmagnetized below the canopy;  this  holds throughout the lower atmosphere. In any case, with the accurate values for collision frequencies presented in Figs.~\ref{fig5}-\ref{fig7}, the actual magnetization and the width of this layer can easily be checked for any given value of the magnetic field.

We can claim   with certainty that there exists a well-defined  (but highly irregular regarding its width) altitude range within which both electrons and protons are totally unmagnetized; this fact should not be ignored in modeling. The mean free path of a particle $j$ (the distance it covers  between two consecutive collisions, given by $\lambda_{fj}=\vtj/\nu_j$) in these  regions is far shorter than the ion gyro-radius. Hence, the  dynamics of ions is not affected by the magnetic field  in most of the photosphere and chromosphere. In some layers this holds for electrons too, and  such an environment can support only waves appropriate for an ordinary gas (e.g. gravity and/or acoustic  oscillations) or heavily damped ion-acoustic waves.

In addition to this,  according to numbers presented above, there exists  an altitude   region within which electrons are magnetized while protons (ions) are not. This makes  it  very difficult to justify  so called two-component models that are  found in recent studies  which assume the medium
to be  composed of neutrals from one side and `plasma' from the other. The term `plasma' here refers  to electrons and ions as a single fluid. Because there are magnetized electrons and unmagnetized ions  in the regions that we clearly identified, we know that the dynamics of the two species  perpendicular  to the magnetic field or at large angles with respect to it   becomes  totally different, which precludes describing   them with a common set of single-fluid equations.

\section{Viscosity and thermal conductivity in unmagneti\-zed plasma}\label{visg}

Because of the altitude-dependent parameters, the  contributions of different species to viscosity and conductivity coefficients will vary in space, and  the spatially dependent contribution of each component should be checked separately. The viscosity tensor components $i, j$ for species $a$ in a strongly collisional plasma-gas mixture  with un-magnetized charged species are given by
\[
\Pi_{a,ij}=-\frac{p_a}{\sum_b \nu_{ab}} \left(\frac{\partial v_{a,i}}{\partial r_j} +  \frac{\partial v_{a,j}}{\partial r_i} - \frac{2}{3} \delta_{ij}
\frac{\partial v_{a,k}}{\partial r_k}\right)
\]
\be
+  \frac{\rho_a}{\sum_b \nu_{ab}} \! \left\{\sum_b \nu_{ab}\left[\left(v_{b, i} - v_{a, i}\right) \left(v_{b, j} - v_{a, j}\right)
- \frac{1}{3} \delta_{ij} \left(\vec v_b - \vec v_a\right)^2\right]\!\right\}.
\label{ve6}
\ee
 Here, $r_{i, j, k}$ stands for  the coordinates $x, y, z$, while  $v_{i, j, k}$ denotes the speed components along these coordinates, and summation (with the general index  $b$) includes all species  including  the specie $a$ itself [i.e., collisions between  alike particle as well, when the terms in the second row  in (\ref{ve6}) clearly vanish].
From Eq.~(\ref{ve6}) it can easily be seen that this  is a symmetric tensor $\Pi_{ij}=\Pi_{ji}$, and its trace is zero, $\Pi_{jj}=0$ (with assumed summation over the repeating index). These are well-known features of the viscosity tensor.

The first row in Eq.~(\ref{ve6}) describes the self-induced viscosity of the species $a$. The second row on the other hand is due to relative motion of the species $a$ with respect to other species (this implies  collisions between dissimilar  particles). This part  may play a key role in the initial stage of some accidental electromagnetic or electrostatic perturbations in which other (uncharged) species are at rest; this holds for the friction force and friction damping as well. Because of collisions between dissimilar  particles, Eq.~(\ref{ve6}) differs considerably from the usual Navier-Stokes formula for single-component gasses.

 The components of the conductivity vector are given by
 \[
 Q_{a, j} = -\frac{5}{3}\frac{p_a}{m_a \sum_b \nu_{ab}}\frac{\partial \kappa T_a}{\partial x_j}
 \]
 \be
  + \frac{\rho_a}{3 \sum_b \nu_{ab}}  \left\{\sum_b \nu_{ab}\left(v_{b, j}- v_{a, j}\right)\left[\left(\vec v_b-\vec v_a\right)^2
     -
  5  \frac{\kappa \left(T_b- T_a\right)}{m_a+ m_b} \right]\right\}.
  \label{ve7}
  \ee
  Similar to the viscosity, here   the term in the first row in Eq.~(\ref{ve7}) is also due to self-conductivity and the remaining terms include interactions  between dissimilar  species.

  Eqs.~(\ref{ve6}), (\ref{ve7}) are obtained from kinetic equation with the Bhatnagar-Gross-Krook (BGK) collisional integral. The Grad method is used together with   the fact that the temperature variation in the photosphere-chromosphere layer studied here is very weak, it changes for about  0.3 eV only. The  general transport coefficients  contained in   Eqs.~(\ref{ve6}), (\ref{ve7})  are given in Sec.~\ref{pd}.

In more explicit form, the components of the viscosity tensor for the un-magnetized species $a$ are
\[
\Pi_{a, xx}=-\frac{p_a}{\sum_b \nu_{ab}}\left(2 \frac{\partial v_{a,x}}{\partial x} - \frac{2}{3} \nabla\cdot \vec v_a\right)
 \]
 \[
 +\frac{m_a n_a}{\sum_b \nu_{ab}}\left\{\sum_b \nu_{ab} \left[\left(v_{b, x} - v_{a, x}\right)^2 - \frac{1}{3}\left(\vec v_b-\vec v_a\right)^2\right]\right\},
\]
\[
\Pi_{a, xy}=-\frac{p_a}{\sum_b \nu_{ab}}\left(\frac{\partial v_{a,x}}{\partial y} +\frac{\partial v_{a,y}}{\partial x} \right)
\]
\[
+ \frac{m_a n_a}{\sum_b \nu_{ab}}\left[\sum_b \nu_{ab} \left(v_{b, x} - v_{a, x}\right)\left(v_{b, y} - v_{a, y}\right)\right]=\Pi_{a, yx},
\]
\[
\Pi_{a, xz}=-\frac{p_a}{\sum_b \nu_{ab}}\left(\frac{\partial v_{a,x}}{\partial z} +\frac{\partial v_{a,z}}{\partial x} \right)
\]
\[
+ \frac{m_a n_a}{\sum_b \nu_{ab}}\left[\sum_b \nu_{ab} \left(v_{b, x} - v_{a, x}\right)\left(v_{b, z} - v_{a, z}\right)\right]=\Pi_{a, zx},
\]
\[
\Pi_{a, yy}=-\frac{p_a}{\sum_b \nu_{ab}}\left(2 \frac{\partial v_{a,y}}{\partial y} - \frac{2}{3} \nabla\cdot \vec v_a\right)
 \]
 \[
 +\frac{m_a n_a}{\sum_b \nu_{ab}}\left\{ \sum_b \nu_{ab} \left[ \left(v_{b, y} - v_{a, y}\right)^2 - \frac{1}{3}\left(\vec v_b-\vec v_a\right)^2\right]\right\},
\]
\[
\Pi_{a, yz}=-\frac{p_a}{\sum_b \nu_{ab}}\left(\frac{\partial v_{a,y}}{\partial z} +\frac{\partial v_{a,z}}{\partial y} \right)
\]
\[
+ \frac{m_a n_a}{\sum_b \nu_{ab}}\left[\sum_b \nu_{ab} \left(v_{b, y} - v_{a, y}\right)\left(v_{b, z} - v_{a, z}\right)\right]=\Pi_{a, zy},
\]
\[
\Pi_{a, zz}=-\frac{p_a}{\sum_b \nu_{ab}}\left(2 \frac{\partial v_{a,z}}{\partial z} - \frac{2}{3} \nabla\cdot \vec v_a\right)
 \]
 \[
 +\frac{m_a n_a}{\sum_b \nu_{ab}}\left\{ \sum_b \nu_{ab}\left[ \left(v_{b, z} - v_{a, z}\right)^2 - \frac{1}{3}\left(\vec v_b-\vec v_a\right)^2\right]\right\}.
\]
 In the expressions above it is appropriate to take $p_a=n_a \kappa T$. Hence, all species have  the same temperature and there is no anisotropy. Both assumptions are well-justified in such a strongly collisional and un-magnetized lower solar atmosphere.

\subsection{Proton dynamics}\label{pd}

Using all previous graphs, we can now calculate the viscosity coefficients  for the particular photospheric plasma. For protons, after  identifying  the leading contributors to their collisions in Sec.~\ref{procol},  we need the following coefficient for the viscosity in the first row of Eq.~(\ref{ve6}):
\be
\eta_{pp}\equiv \frac{n_p(x)  \kappa T(x)}{\nu_{pp}(x)+ \nu_{p{\sss H}}(x)+ \nu_{p{\sss H}e}(x) + \nu_{p{\sss H}e^+}(x)}, \quad \mbox{[in kg/(s\,m)]}.\label{vis1}
\ee
In the second row of Eq.~(\ref{ve6}) we have the following four coefficients [all in units kg/m$^3$]:
\be
\mu_{pp}\equiv \frac{m_p n_p(x) \nu_{pp}(x) }{\nu_{pp}(x)+ \nu_{p{\sss H}}(x) + \nu_{p{\sss H}e}(x) + \nu_{p{\sss H}e^+}(x)}, \label{vis2}
 \ee
 \be
 \mu_{p{\sss H}}\equiv \frac{m_p n_p(x) \nu_{p{\sss H}}(x) }{\nu_{pp}(x)+ \nu_{p{\sss H}}(x)+ \nu_{p{\sss H}e}(x)  + \nu_{p{\sss H}e^+}(x)},\label{vis3}
\ee
\be
\mu_{p{\sss H}e}\equiv \frac{m_p n_p(x) \nu_{p{\sss H}e}(x) }{\nu_{pp}(x)+ \nu_{p{\sss H}}(x)+ \nu_{p{\sss H}e}(x)  + \nu_{p{\sss H}e^+}(x)},\label{vis4}
\ee
 \be
\mu_{p{\sss H}e^+}\equiv \frac{m_p n_p(x) \nu_{p{\sss H}e^+}(x) }{\nu_{pp}(x)+ \nu_{p{\sss H}}(x) + \nu_{p{\sss H}e}(x) + \nu_{p{\sss H}e^+}(x)}.\label{vis5}
\ee
Note that to calculate $\mu_{p{\sss H}}(x)$ and $\mu_{p{\sss H}e}(x)$ we have to use the viscosity lines (line 4 from Fig.~\ref{fig1} and line 3 from Fig.~\ref{fig2}, respectively) and the top $x$-axis for the energy.

\begin{figure}[!htb]
   \centering
\includegraphics[height=6cm,bb=17 14 274 215,clip=]{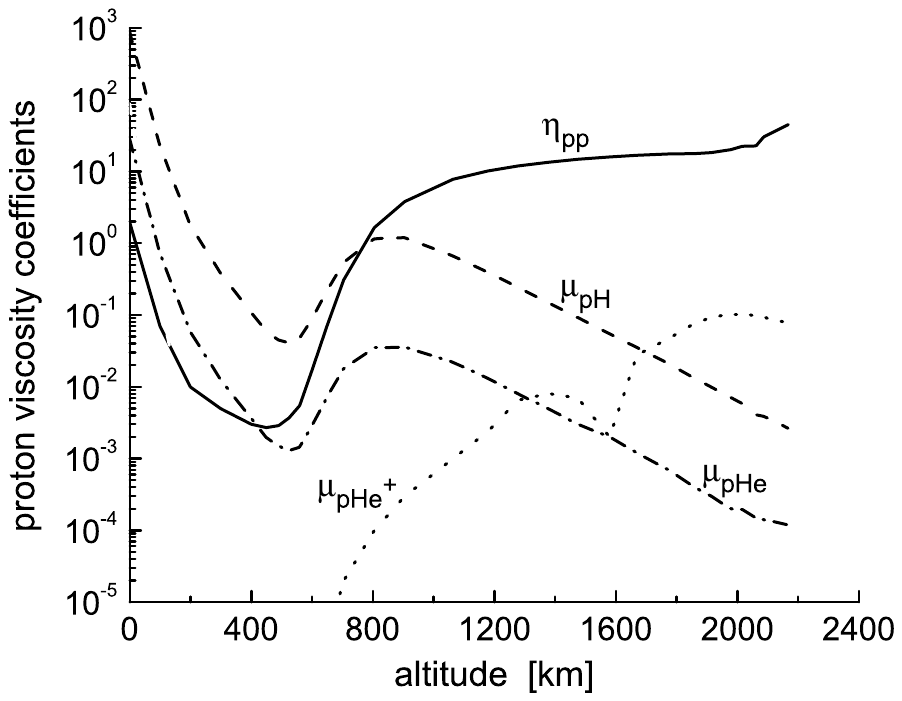}
\caption{Dynamic viscosity coefficients (\ref{vis1}), (\ref{vis3}-\ref{vis5}) without the magnetic field for protons in the lower solar atmosphere.   Here $\eta_{pp}$ is plotted in units $10^{-8}$ kg/(s m) and all other coefficients in  $10^{-10}$ kg/m$^3$.    \label{fig11a}}
\end{figure}

The numerical  values for the coefficients (\ref{vis1}) and (\ref{vis3}-\ref{vis5}) are  presented in Fig.~\ref{fig11a}.
The coefficient $\mu_{pp}$ is not presented  because from the second row in Eq.~(\ref{ve6}) it is seen that for $a=b$ its contribution vanishes. Therefore the most dominant $\mu_{ab}$ terms should be checked only for the case $a\neq b$. From Fig.~\ref{fig11a} it is clear that  in most of the space  the coefficient  $\mu_{p{\sss H}}$  should be taken into account. However, above an  altitude of about  1700 km the viscosity between protons and helium ions becomes the most dominant, as  is seen from the dotted line,  which gives the values of $\mu_{p{\sss H}e^+}$, though this  may change if protons are magnetized; then  the  gyro-viscosity should be taken into account. This  problem   will be discussed elsewhere. Furthermore  the viscosity that  involves neutral helium $\mu_{p{\sss H}e}$ is clearly negligible everywhere.

It is meaningless to directly  compare the leading $\mu_{ab}$ term with $\eta_{pp}$    because they are in  different units. Instead, one must compare the complete corresponding viscosity terms from the first and second row  in Eq.~(\ref{ve6}), which is approximately
\be
R_{\eta \mu}\simeq\frac{\eta_{pp} \frac{\partial \displaystyle{v_{p,i}}}{\partial \displaystyle{r}}}{\mu_{p{\sss H}} \delta v_i \delta v_j}
\simeq \frac{\eta_{pp}}{\mu_{p{\sss H}}} \frac{v_{p,i}}{L_v  \delta v_i \delta v_j }, \quad \delta v_i= v_{b, i}- v_{p, i}. \label{rat}
\ee
For waves with the wave number k we have $L_v\simeq 1/k$. Because the ratio  $\eta_{pp}/\mu_{p{\sss H}}$ changes with altitude for many orders of the magnitude, the relative contribution of the two terms to the viscosity will drastically change. For example, at $x=0$  we have $\eta_{pp}/\mu_{p{\sss H}}=0.22$ [m$^2$/s] and at
$x=1580$ km $\eta_{pp}/\mu_{p{\sss H}}=3.2 \cdot 10^4$ [m$^2$/s]. The speed difference $\delta v_i$ may be  time dependent, for instance for waves that first affect  some charged species $a$ while the un-charged species $b$ is set into motion only after some collisional time, which  will also affect the ratio (\ref{rat}).

Without  relative motion between protons and other species,  the proton viscosity is strictly self-induced. With  relative motion the situation becomes much more complicated and it is not so obvious which terms are more dominant. This is true in particular in view of ratio (\ref{ratvf}), which involves friction (see below). Hence, to be on a safe ground, one should keep the two leading viscosity terms discussed above  together with the corresponding friction force terms.

\begin{figure}[!htb]
   \centering
\includegraphics[height=6cm,bb=17 15 271 208,clip=]{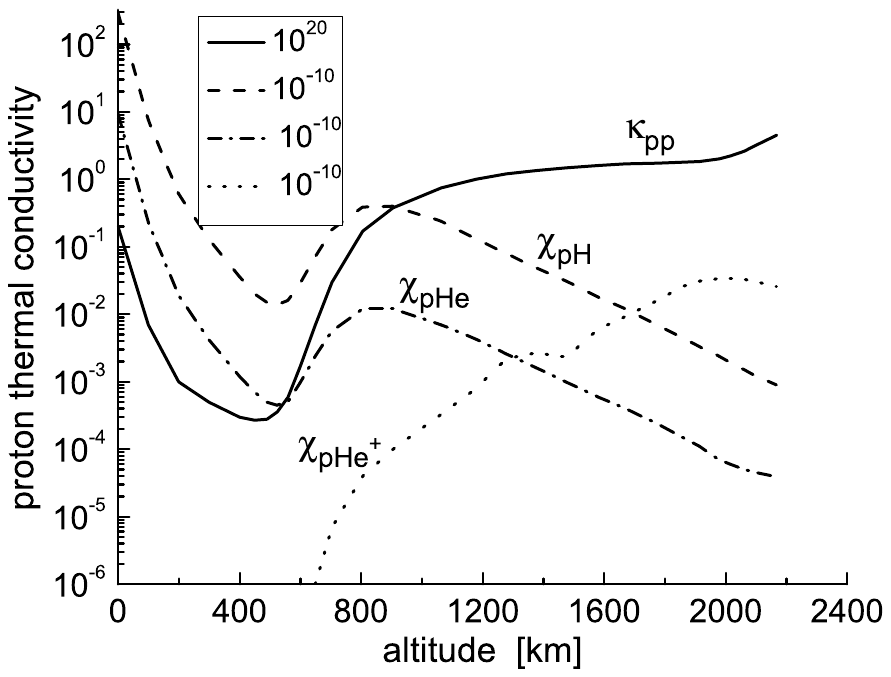}
\caption{Proton thermal conductivity coefficients (\ref{kpp}-\ref{kajhei}) with altitude; $\kappa_{pp}$ is plotted in (s\,m)$^{-1}$, and  $\chi_{pj}$  in kg/m$^3$. \label{fig11}}
\end{figure}

A similar analysis is now used for the conductivity vector (\ref{ve7}). For unperturbed situations the second row  in Eq.~(\ref{ve7}) can usually be neglected because i) collisions couple separate species and they are expected to move together, and ii) in the presence of frequent collisions thermalization is very effective, so the temperatures are equal. This
does not necessarily hold for  some accidental electrostatic or electromagnetic  perturbations. For example,  electrostatic ion-acoustic-type perturbations in the given environment must involve ion temperature perturbations (because the wave phase speed is on the same order as the ion thermal speed), and this happens on the background of initially static neutrals. Therefore the second row  in (\ref{ve7}) should be kept. For protons, keeping the most relevant terms as above, we have the $x$-component of the conductivity vector
\[
Q_{px}\!=-\kappa_{pp} \frac{\partial}{\partial x}\left(\kappa T\right)
+ \chi_{p\sss{H}} \left(v_{\sss{H},x}-v_{p,x}\right) \left[\!\left(\vec v_{\sss{H}}-\vec v_p\right)^2\!- \!5 \frac{\kappa \left(\!T_{\sss{H}}\! - \!T_p\!\right)}{m_{\sss{H}}\! + \!m_p}\!\right]
\]
\[
+ \chi_{p\sss{H}e}\left(v_{\sss{H}e,x}-v_{p,x}\right)\left[ \left(\vec v_{\sss{H}e}-\vec v_p\right)^2
 -5 \frac{\kappa \left(T_{\sss{H}e} - T_p\right)}{m_{\sss{H}e} + m_p}\right]
 \]
 \[
  +  \chi_{p\sss{H}e^+}\left(v_{\sss{H}e^+,x}-v_{p,x}\right)\left[ \left(\vec v_{\sss{H}e^+}-\vec v_p\right)^2
 -5 \frac{\kappa \left(T_{\sss{H}e^+} - T_p\right)}{m_{\sss{H}e^+} + m_p}\right].
\]
In view of Eq.~(\ref{ve7}), the thermal conductivity  coefficients which appear here  are
\be
\kappa_{pp}=\frac{5}{3} \frac{n_p v_{{\sss T}p}^2}{\nu_{pp}+ \nu_{p{\sss H}} + \nu_{p{\sss H}e} + \nu_{p{\sss H}e^+}}, \quad \mbox{$ \left[\frac{1}{s m}\right]$}, \label{kpp}
 \ee
 \be
 \chi_{p\sss{H}}=\frac{1}{3}\frac{m_p n_p \nu_{p{\sss H}}}{\nu_{pp}+ \nu_{p{\sss H}} + \nu_{p{\sss H}e} + \nu_{p{\sss H}e^+}}, \quad \mbox{$ \left[\frac{kg}{m^3}\right]$}, \label{kajh}
\ee
\be
\chi_{p\sss{H}e}=\frac{1}{3}\frac{m_p n_p \nu_{p{\sss H}e}}{\nu_{pp}+ \nu_{p{\sss H}} + \nu_{p{\sss H}e} + \nu_{p{\sss H}e^+}}, \label{kajhe}
\ee
\be
\chi_{p\sss{H}e^+}=\frac{1}{3}\frac{m_p n_p \nu_{p{\sss H}e^+}}{\nu_{pp}+ \nu_{p{\sss H}} + \nu_{p{\sss H}e} + \nu_{p{\sss H}e^+}}.\label{kajhei}
\ee
Here, similar to Eqs.~(\ref{vis1}-\ref{vis5}), all parameters are altitude dependent, which affects the thermal conductivity coefficients (\ref{kpp}-\ref{kajhei}), whose altitude dependence is  presented in Fig.~\ref{fig11}  in units $10^{20}$ (s\,m)$^{-1}$ for $\kappa_{pp}$,  and $10^{-10}$ kg/m$^3$ for $\chi_{pb}$ coefficients. Similar to viscosity, regarding $\chi_{pb}$ coefficients,  here again proton interaction with hydrogen atoms is most dominant below 1700 km and only $\chi_{p\sss{H}}$ should be kept in the part that  describes the interaction between different species in Eq.~(\ref{ve7}). Above this altitude one should keep  $\chi_{p\sss{H}e^+}$ only. These conclusions   hold as long as the ions are un-magnetized (gyro-viscosity  excluded). Identifying these most dominant terms can significantly simplify the derivations.

\subsection{Friction vs. viscosity. Vanishing friction}

Comparing the contribution to the momentum equation of the  viscosity term from the second row in Eq.~(\ref{ve6}) and the usual friction force  yields the following approximate dimensionless ratio of these  terms for the species $a$:
\be
R_{1,a}=\frac{F_v}{F_f}\simeq \frac{\sigma_{v,ab}}{\sigma_{m, ab}} \frac{k (v_a-v_b)}{\sum_b \nu_{a}}. \label{ratvf}
\ee
Here, we estimate the interaction between the species $a$ and another single species $b$, and $\sigma_{v,ab}/\sigma_{m, ab}<1$ is the ratio of the cross sections for viscosity and momentum transfer [which are typically different, see Figs.~\ref{fig1}-\ref{fig3}], and $k^{-1}\equiv \lambda$ is the characteristic scale length for the speed difference (e.g., wave length for wave analysis), which appears from $\nabla\cdot \Pi_{a,ij}$ in the momentum equation.  It is clear that $ R_{1,a}$ can have any value [e.g., because the difference $v_a-v_b$ may be time dependent for  electromagnetic or electrostatic wave phenomena that  affect the charged species $a$ first]. Therefore it cannot be justified to neglect   the second row in Eq.~(\ref{ve6}) in the wave analysis. Assuming that $a$ represents  charged species and $b$ some neutral one, for  electromagnetic or electrostatic perturbations  species $b$ is initially at rest so that $R_{1,a}<1$ if
\[
\lambda>\frac{v_a}{\sum_b \nu_{ab}}\frac{\sigma_{v,ab}} {\sigma_{m, ab}}.
\]
For these wavelengths the contribution of the second row in the viscosity  in Eq.~(\ref{ve6}) is negligible, but  this holds in principle  for the initial regime  only.

Comparing now the viscosity contribution to the momentum equation due to the first row in Eq.~(\ref{ve6}), with the friction force between $a$ and $b$ for the initial stage with $v_b=0$, yields
 \be
R_{2,a}=\frac{k^2 \vta^2}{\nu_{m, ab} \sum_b \nu_{v,a}}. \label{ratvf2}
\ee
The friction force dominates if
\[
\lambda^2>\frac{\vta^2}{ \nu_{m, ab} \sum_b \nu_{v,a}}.
\]
The index $m$ here denotes the collision frequency calculated with the appropriate cross section for momentum transfer. Hence, the friction force is stronger provided that the wavelength exceeds both of these expressions, i.e.,
\be
\lambda> \frac{v_a}{\sum_b \nu_{ab}}\frac{\sigma_{v,ab}} {\sigma_{m, ab}}, \quad \frac{\vta}{\left( \nu_{m, ab} \sum_b \nu_{v,a}\right)^{1/2}}.\label{wl}
\ee
The presented conclusions  hold for the initial regime only. The species are coupled through collisions and the speed difference $v_a-v_b$, in both the friction force and in the viscosity terms containing the speed difference may approach  zero provided the time interval is long enough. This issue (related to friction force) has been discussed  in Vranjes et al. \cite{v3}. The velocity difference relaxation is altitude dependent, and after presenting the  detailed collision frequencies in the previous sections, we can make some estimates for proton interaction with neutrals. Assuming that protons start to move with some initial speed $v_{p0}$ through static background of hydrogen atoms, we can calculate the time needed for both species to achieve some common speed $v_c$, which naturally must be between $0$ and $v_{p0}$. This is performed  for several layers to see the differences caused by  density and temperature variation.
Starting from the momentum equations with friction only $\partial \vec v_n/\partial t= \nu_{np} (\vec v_p - \vec v_n)$, $\partial \vec v_p/\partial t= \nu_{pn} (\vec v_n - \vec v_p)$, we find  the time dependent velocity of the two species:
 \be
\vec v_n=v_c +
\frac{\left(\vec v_{n0} - \vec v_{p0}\right) \nu_{np}}{\nu_{pn}+ \nu_{np}}
\cdot\exp[-(\nu_{pn} + \nu_{np})t], \label{pari} \ee
\be
 \vec v_p=v_c -
\frac{\left(\vec v_{n0} -\vec v_{p0}\right) \nu_{pn}}{\nu_{pn}+ \nu_{np}}
\cdot\exp[-(\nu_{pn} + \nu_{np})t]. \label{parn} \ee
 Taking $m_p\sim m_n$ and  $\vec v_{n0}=0$, we obtain for the common velocity for both species
 \be
v_c \equiv \frac{\nu_{pn} \vec v_{n0} + \nu_{np} \vec v_{p0}}{\nu_{pn}+ \nu_{np}}=v_{p0}\frac{ \nu_{np}}{\nu_{pn} + \nu_{np}} = v_{p0} \frac{n_{p0}}{n_{p0}+
n_{n0}}.\label{com} \ee
%
%
\begin{table}
\caption{Common velocity and  velocity relaxation time $t_c$  for protons (with arbitrary starting speed $v_{p0}$) and hydrogen (with $v_{n0}=0$) for several altitudes.   }             
\label{table3}      
\centering                          
\begin{tabular}{lllll }        
\hline\hline                 
$x$ [km] & $0$ & $650$ & $1180$ & $2043$ \\    
\hline                        
  $v_c$ & $5.1\cdot 10^{-4} v_{p0}$ & $2.2\cdot 10^{-5} v_{p0}$ & $9.5\cdot 10^{-3} v_{p0}$ & $0.37 v_{p0}$\\      
  $t_c$ [s] & $5.6\cdot 10^{-9}$ & $1.2\cdot 10^{-6}$ & $7.6\cdot 10^{-5}$ & $5\cdot 10^{-3}$ \\    
\hline                                   
\end{tabular}
\end{table}
In Table~\ref{table3} the common speed (\ref{com}) is given for several altitudes, and the time necessary for both species to achieve 99 percent of this common speed is calculated from
\[
t_c=-\frac{\ln[0.01]}{\nu_{pn} \left(1+ n_p/n_n\right)}.
\]
To calculate $\nu_{pn}$, which appears here, we used the cross section for momentum transfer from Fig.~\ref{fig1}.

From Table~\ref{table3} we can conclude the following.  a) Because of the very frequent collisions between the two species, the friction  between them should vanish very quickly, but b) at  the same time this must have a strong effect on the small electrostatic or electromagnetic perturbations that are now supposed to set into motion the two species. This implies c) that the amplitude of these  perturbations must reduce dramatically. d) Because the two species move together (after the collisional time $t_c$), the friction is effectively zero, therefore the only relevant remaining dissipation mechanism must be through viscosity. e) All these conclusions are altitude dependent; higher up the friction may become dominant dissipation process, in particular for a certain wavelength regime, as predicted by  Eq.~(\ref{wl}).

These facts are frequently overlooked in the literature where  such a regime of common motion is just assumed [but without much regard to the consequence c) above] and  equations for different components are summed up, and a single fluid dynamics is then studied. This implies that the transition process, that takes place within the collision time, is neglected together with the physical phenomena involved in the process. This problem is discussed in Vranjes et al. \cite{v3}. A statement of the validity of common equations for all species is given also in  Alfv\'{e}n  and F\"{a}lthammar \cite{alf} on p. 177, where the authors write that the common speed has sense only if the speed of neutrals is nearly equal to the speed of plasma. Such a situation  surely cannot be expected in the initial stadium  of some electrostatic or electromagnetic perturbations that naturally affect plasma species first, with the background of initially immobile neutrals whose dynamics develops  due to friction  and partly due to viscosity, and  this after the collisional time only [Vranjes et al. \cite{v3}]. In the present work we are able to quantify these phenomena by calculating the precise characteristic collisional time  and consequently by predicting the amplitude of the common speed achieved within such a time interval.

\subsection{Hydrogen dynamics}\label{hd}
\begin{figure}[!htb]
   \centering
\includegraphics[height=6cm,bb=17 15 275 213,clip=]{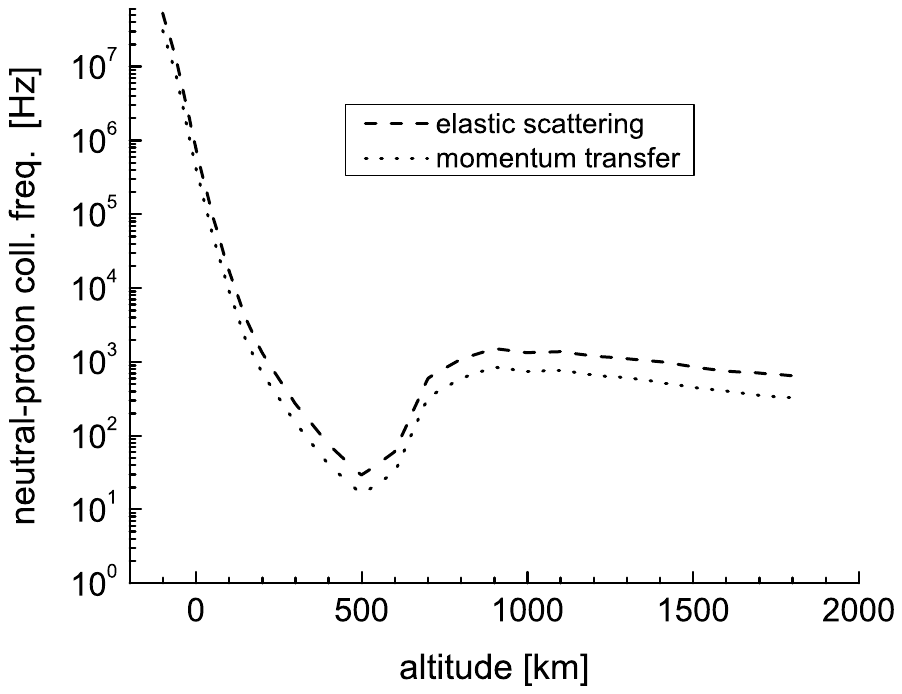}
\caption{Collision frequency for $H-p$ collisions. \label{neut-p}}
\end{figure}

We have seen that proton interaction with neutral hydrogen is the most dominant in most of the space, so  in a proper wave analysis that obeys  conservation laws, one needs  the continuity, momentum, and energy equations describing hydrogen dynamics as well. Hydrogen collision frequencies with other species (electrons, ions, and helium atoms) and corresponding cross sections  can easily be obtained using the momentum conservation
 \be
 m_j n_j \nu_{j{\sss H}}=m_{\sss{H}} n_{\sss{H}} \nu_{{\sss H}j}. \label{mc}
 \ee
 For completeness, the collision frequency $\nu_{{\sss H} p}$ for the most dominant neutral hydrogen atom collisions with  protons is presented in Fig.~\ref{neut-p} for the elastic scattering (dashed line) and momentum transfer (dotted line). The local minimum in the profile is due to the decrease in  the profile of target particles (protons).

We furthermore need components for  hydrogen viscosity and thermal conductivity.
For $\eta_{\sss HH}, \kappa_{\sss HH} $ we may set
\be
\eta_{\sss HH}=\frac{n_{\sss H} \kappa T}{\sum_b \nu_{{\sss H}b}}\simeq \frac{n_{\sss H} \kappa T}{\nu_{{\sss H H}}}, \quad
\kappa_{\sss HH}=\frac{5}{3} \frac{n_{\sss H} v_{t{\sss H}}^2}{\sum_b \nu_{{\sss H}b}}\simeq \frac{5}{3} \frac{n_{\sss H} v_{t{\sss H}}^2}{\nu_{\sss H H}}.\label{ekbgk}
\ee
With the cross section $\sigma_{{\sss HH}}$ determined by  line 2 in Fig.~\ref{fig3}, the dynamic viscosity coefficient for hydrogen self-collisions becomes
\be
\eta_{\sss HH}=\frac{m_{{\sss H}} v_{{\sss T}}}{\sigma_{{\sss HH}}}. \label{ev}
\ee
 In writing it we used $\nu_{{\sss HH}}= \sigma_{{\sss HH}} n_{{\sss H}} v_{{\sss T}}$.

In the literature one can find also the expression that  follows from the Chapman and Cowling (\cite{cc}) model  based on the interaction of  hard spheres:
\be
\eta_{cc} = \alpha m n_{{\sss H}} \nu_{{\sss HH}} \lambda_f^2, \quad \alpha=\frac{75 \sqrt{\pi}}{64} \left(1+ \frac{3}{202}+ \cdots\right), \label{ev2}
\ee
where  $\lambda_f=1/(\sqrt{2} \pi n_{{\sss H}} r_{{\sss H}}^2)$ is the mean free path, and  $r_{{\sss H}}$ is the diameter of the colliding particles, in the present case its value is $r_{{\sss H}}=2.12 \cdot 10^{-10}$ m. This yields
 \be
\eta_{cc}=\frac{0.47 m  v_{{\sss T}}}{r_{{\sss H}}^2}. \label{ve3}
\ee
The expressions (\ref{ev}) and (\ref{ve3}) are checked against experimental measurements available in Vargaftik et al. \cite{varg} for a hydrogen gas, and the results for several temperatures  are presented in Table~\ref{tab-2}.   In the given energy range our $\eta_{\sss HH}$ gives values closer to the experimental ones. The  differences between the two  models are  about factor 2, which is clearly due to indistinguishability effect that is missing in the Chapman and Cowling classical model.
%
%
\begin{table}
\caption{Hydrogen dynamic viscosity coefficient [in units kg/(sm)= Pa$\cdot$s] for several temperatures. Second row: temperature-dependent cross section from Fig.~\ref{fig3}. Third  row: our value based on Eq.~(\ref{ev}). Fourth row: value based on the Chapman and Cowling hard sphere model (\ref{ev2}). Fifth row: experimental values for pure neutral hydrogen from Vargaftik et al. \cite{varg}.  }
\label{tab-2}      
\centering                          
\begin{tabular}{lllll}        
\hline\hline                 
$T$ [K]& $4400$ & $4990$ & $6560$& $11150$  \\    
\hline                        
 $\sigma_{{\sss HH}}$ (a.u.) & $45.35$ & $45.3$ & $40$ & $ 33.5$ \\      
 $\eta_{\sss HH}$ & $3.95\cdot 10^{-5}$ & $4.2\cdot 10^{-5}$    & $5.5\cdot 10^{-5}$ & $0.86\cdot 10^{-4}$ \\
   $\eta_{cc}$ & $ 10 \cdot 10^{-5}$ & $11\cdot 10^{-5}$     & $13\cdot 10^{-5}$ & $1.67\cdot 10^{-4}$\\
  $\eta_{exp}$ & $4.8\cdot 10^{-5}$ & $5.5\cdot 10^{-5}$    & $7\cdot 10^{-5}$ & $1.1\cdot 10^{-4}$\\
  \hline                                   
\end{tabular}
\end{table}

Using the Chapman and Cowling \cite{cc} model, we can also calculate  the coefficient of thermal conductivity for hydrogen gas
\[
\kappa_{cc}=n_{{\sss H}} \frac{5 \sqrt{\pi}}{16} \left(1+ \frac{1}{44}+\cdots\right) \nu_{{\sss HH}} \lambda_f^2\simeq
\frac{5}{16\sqrt{2 \pi}}  \left(1+ \frac{1}{44}\right)\frac{ v_{{\sss T}}}{r_{{\sss H}}^2}
\]
\be
=2.84\cdot 10^{18} v_{{\sss T}}, \quad \mbox{[in $(sm)^{-1}$]}. \label{ve4}
\ee
This can  be compared with the above-given corresponding coefficient (\ref{ekbgk}) obtained from kinetic theory with the BGK collision integral:
\be
\kappa_{\sss HH}=\frac{5}{3}\frac{p_{{\sss H}}}{m_{{\sss H}} \nu_{{\sss HH}}}= \frac{5}{3}\frac{ v_{{\sss T}}}{\sigma_{{\sss HH}}}. \label{ve5}
\ee
 Around the temperature minimum  region in the photosphere, using line 2 from Fig.~\ref{fig3} for hydrogen,  our conductivity coefficient
$\kappa_{\sss HH}$ is higher by about a factor 2 than the Chapman and Cowling coefficient (\ref{ve4}), which is the consequence of the quantum-mechanical indistinguishability incorporated in our derivations.

For the viscosity coefficients $\mu_{{\sss H} b}$, which are associated with the terms containing the speed difference between $H$ and $b$ species, the situation is as follows:  The speed difference between different {\em neutral} species cannot be of any importance for  obvious reasons (they are coupled through collisions, they react similarly to perturbations by external forces). Hence, both $\mu_{\sss HH }$ and  $\mu_{{\sss H H}e }$ are not needed for the same reason. We now compare
\[
 \frac{\mu_{{\sss H }p }}{ \mu_{{\sss H H}e^+ }}=\frac{n_p}{n_{{\sss H}e^+}} \frac{\sigma_{{\sss H} p}}{\sigma_{{\sss HH}e^+}}.
 \]
From Fontenla et al. \cite{fon} we know that $n_p\gg n_{{\sss H}e^+}$ in the whole region of interest here. Therefore  $\mu_{{\sss H H}e^+ }$ is most likely
negligible. Using (\ref{mc}) we have
\be
\mu_{{\sss H }p }\simeq \frac{m_{\sss H} n_{\sss H} \nu_{{\sss H} p}}{\nu_{{\sss H H}}}= \frac{m_{\sss H} n_p \sigma_{p  {\sss H}}}{ \sigma_{\sss H H}}.\label{muhp}
\ee
Similar arguments are used for the coefficients $\chi_{{\sss H} b}$ in the conductivity vector. Hence, the only remaining coefficient we need is
\be
 \chi_{{\sss H} p}\simeq\frac{m_{\sss H}  n_{\sss H } \nu_{{\sss H}p}  }{3 \nu_{{\sss HH}}}= \frac{m_{\sss H}  n_p \sigma_{p{\sss H}}  }{3 \sigma_{{\sss HH}}}.\label{chihp}
 \ee
  \begin{figure}[!htb]
   \centering
\includegraphics[height=6cm,bb=17 14 270 211,clip=]{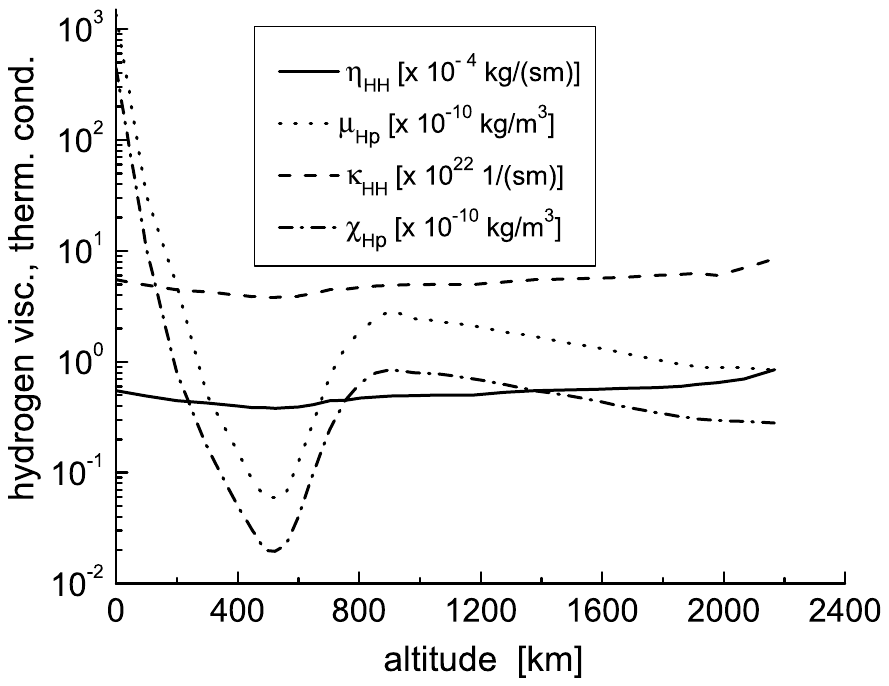}
\caption{Most relevant viscosity ($\eta_{\sss HH}, \mu_{{\sss H }p }$)  and thermal conductivity ($\kappa_{\sss HH}, \chi_{{\sss H} p}$) coefficients for hydrogen atoms. \label{fig12}}
\end{figure}

The most important coefficients for hydrogen $\eta_{\sss HH}, \mu_{{\sss H }p }, \kappa_{\sss HH}, \chi_{{\sss H} p}$  are presented in
Fig.~\ref{fig12}. Evidently,  for practical purposes in the lower solar atmosphere the self-interaction coefficients $\eta_{\sss HH}, \kappa_{\sss HH}$ may be taken as constant and their values are  $\simeq 0.5\cdot 10^{-4}$ kg/(sm) and $\simeq 5\cdot 10^{23}$ (sm)$^{-1}$, respectively.  The other two coefficients $\mu_{{\sss H }p }$ and $\chi_{{\sss H} p}$, which include interaction with other species are strongly altitude dependent.

\section{Summary and discussions}\label{sum}

The parameters in the lower solar atmosphere change with  altitude and much  care is needed to properly describe the physical processes that take place there. One  obvious example is presented in Fig.~\ref{fig5}  where the  e-H collision frequency  changes by seven orders of magnitude between the altitudes of $-100$ km and $2200$ km, taking values  $1.3\cdot10^{10}$ Hz and $2.4\cdot 10^3$ Hz, respectively. In addition to this, the type  of collisions changes as well,  e-H collisions being the most dominant up to 900 km and  e-p collisions  above that.   A detailed knowledge  of these  processes is  essential to estimate the friction and related phenomena (e.g. conductivity, transport, etc.).

Our  most important conclusions can be  summarized as follows:
\begin{description}
\item{i)} The cross sections presented in Sec.~2.1 are the most accurate existing ones. They contain the following essential details: a)  variation of cross sections with temperature (altitude), b) variation of cross sections due to quantum effects in the given low-temperature range in the lower solar atmosphere, and c) clear and pronounced differences of cross sections describing  elastic scattering, momentum transfer, and viscosity. Combined, these  fine details may introduce significant differences for various processes related to magnetization, transport, heating, etc.  Although well-known in the laboratory plasmas, so far these  details have not been studied in the solar atmosphere.
     Unlike the various approximate data,   the data we used here (all of them from cited references) are fully quantum-mechanical and  obtained without (almost) any approximation, thus are  the most accurate data ever obtained in  collisional physics (numerical accuracy to six significant digits, physical accuracy bellow one percent). The  only assumption used in their derivation was that the electronic excitation to the excited nonresonant electronic states is negligible (though charge transfer is included). This approximation is quite accurate below the energy threshold for electronic excitation and this is the source of the physical accuracy of `only' one percent.  In deriving  these data even ro-vibrational degrees of freedom were taken, and they are created to serve as a benchmark for checking the accuracy of  other approximate approaches. Needless to say, we  used the exact ion-atom potentials  from  $R=0$ to $R=80000$ a.u.

  The cross sections presented here, coincide with classical at high energies. At low energies (roughly below 1 eV) the cross sections include the effects of a) indistinguishability and b) charge transfer. The lower solar atmosphere is indeed within this low-energy range, and consequently these  intrinsic properties of the plasma-gas matter  cannot be avoided. The accuracy of our collision data for ion-atom collisions removes possible  doubts on the size of the momentum exchange used in previous works available in the literature, including possible under- or overestimation of the role of Alf\'{e}n waves and kink wave damping in the lower solar atmosphere.

\item{ii)} For electron dynamics above 850 km neutrals plays no  practical role, although the neutral number density at 850 km is still three orders of magnitude higher than that of protons and electrons. However, this may not be so if inelastic collisions are taken into account, e.g., those in which electrons are lost or created, as discussed in Sec.~2. These phenomena are beyond the scope of the present work.

\item{iii)} For proton dynamics the role of neutrals is negligible above 1900 km, although  at this altitude (according to data from Fontenla et al. \cite{fon})  $n_p=4.24\cdot 10^{16}$ m$^{-3}$ is still below the neutral hydrogen density  $n_{{\sss H}}=1.7\cdot 10^{17}$ m$^{-3}$.   We stress again that this conclusion may not hold if  inelastic collisions are taken into account.

\item{iv)} There exists a    layer    within which both electrons and ions are definitely un-magnetized.  For intense magnetic structures with a magnetic field of $0.1$ T this layer is located around an  altitude $x=0$ and below.  In this layer the magnetic field plays no direct role in the dynamics of both electrons and ions.

\item{v)} The layer of unmagnetized electrons and ions  continues with a  much thicker  layer  in which electrons are magnetized and ions are not.  The depth of this layer  changes spatially, and  for protons it  is at least 1000 km thick  in regions with a kilo-Gauss magnetic field.  The dynamics of electrons and ions  in this region is completely different, and models that  assume a two-component system consisting of `neutrals' on  one side and `plasma' on the other are meaningless. This is because the `plasma' contains electrons and ions whose dynamics is totally different because of  the magnetic field. Consequently, in this layer the electrons and ions cannot be treated as a single fluid. A fully multi-component analysis (fluid or kinetic) has no alternative in the lower solar atmosphere.

    \item{vi)} Viscosity and thermal conductivity coefficients given in this work are currently  the most accurate, and at  the same time the most complete ones because they contain all most relevant terms appropriate for a multi-component system such as is the lower solar atmosphere  with un-magnetized ions. They  also completely agree with experimental measurements. Our results show that including viscosity both for protons and neutral hydrogen  may be  essential  to properly capture diffusion phenomena in the solar atmosphere.
\end{description}

If we return back to Fig.~\ref{fig1}, we see that the proton-hydrogen cross section changes with temperature (i.e. with the altitude). Its values for elastic scattering (in the laboratory frame) are $2.269\cdot 10^{-18}$ m$^{2}$ and $1.601\cdot 10^{-18}$ m$^{2}$ at temperatures of $5\cdot 10^3$ K and $20 \cdot 10^3$ K, respectively. The temperature of  $5\cdot 10^3$ K corresponds to the altitude 200 km (and also 705 km), and $20 \cdot 10^3$ K corresponds to the altitude $\simeq 2200$ km. We can can now compare these with the cross section used by  other researchers, e.g.,  Zaqarashvili et al. \cite{zak}, where the cross section was assumed to be  constant with the value $8.79\cdot 10^{-21}$ m$^{2}$, i.e., $\pi$ a.u.,  describing collisions where ions and neutrals are treated as hard spheres.  Our  correct values, which include the quantum-mechanical effect of indistinguishability,  for the two temperatures given above are 258  and 182 times greater! We note that these authors make no distinction between cross sections for elastic scattering and momentum transfer. Therefore we can also compare their value with our  cross section for the momentum transfer from the same figure; for the two temperatures we have $1.040\cdot 10^{-18}$ m$^{2}$ and $8.785\cdot 10^{-19}$ m$^{-2}$. These are again  118 and 100 times greater  than their values.
We observe that in the mentioned work  the proton-helium cross section  is assumed to be the same as the one for proton-hydrogen collisions given above. However, from our Fig.~\ref{fig2} the cross sections for p-He elastic scattering at $5\cdot 10^3$ K and $20 \cdot 10^3$ K in the laboratory frame  are about 110  and 70 times  greater than their value. At the same time, our cross section for momentum transfer for the two temperatures is  20 and 8 times greater  than their value.

On the other hand, the cross section for p-H collisions in Khomenko and Collados \cite{kho} is fixed to
$5\cdot10^{-19}$ m$^2$, which is rather close to our value for the momentum transfer cross section in plasma reference frame, roughly speaking only twice as  smaller.  Though in their subsequent  calculation of the collision frequency this difference is  compensated by the numerical factor $8^{1/2}$, which they keep in the thermal velocity, and the collision frequency which they obtain is very close to our value.

In view of the results presented here a natural next step  is to include effects of inelastic collisions.  In our previous work (Vranjes and Poedts \cite{vpla})  we have shown that in certain layers in the photosphere all ions in a unit volume recombine at least 26 times per second. This may have consequences  on magnetization, for example. Perhaps this may be used also to explain the nature and longevity of prominences, which are believed to contain considerable amounts of neutrals. Their longevity  is a challenge for the theory because neutrals should naturally  diffuse and evacuate a prominence by moving to lower layers due to gravity. However, in the presence of inelastic collisions this  diffusion should take place at reduced speed because a neutral  particle does  not remain neutral all the time, it is consequently affected by the magnetic field and the prominence may last longer.

\end{document}